\documentclass[twocolumn,showpacs,preprintnumbers,amsmath,amssymb]{revtex4}

\usepackage{graphicx}
\usepackage{dcolumn}
\usepackage{bm}

\begin{document}

\preprint{IUHET-490}

\title{Interacting Dark Energy and the Cosmic Coincidence Problem}

\author{Micheal S. Berger}
 \email{berger@indiana.edu}
\author{Hamed Shojaei}%
 \email{seshojae@indiana.edu}
\affiliation{%
Physics Department, Indiana University, Bloomington, IN 47405, USA}

\date{\today}

\begin{abstract}
The introduction of an interaction for dark energy to the standard 
cosmology offers a potential solution 
to the cosmic coincidence problem. We examine the conditions on 
the dark energy density that must be satisfied for this scenario to be 
realized. Under some general conditions we find a stable attractor for the 
evolution of the Universe in the future. Holographic conjectures for the 
dark energy offer some
specific examples of models with the desired properties.
\end{abstract}

\maketitle

\def\al{\alpha}
\def\be{\beta}
\def\ga{\gamma}
\def\de{\delta}
\def\ep{\epsilon}
\def\ve{\varepsilon}
\def\ze{\zeta}
\def\et{\eta}
\def\th{\theta}
\def\vt{\vartheta}
\def\io{\iota}
\def\ka{\kappa}
\def\la{\lambda}
\def\vpi{\varpi}
\def\rh{\rho}
\def\vr{\varrho}
\def\si{\sigma}
\def\vs{\varsigma}
\def\ta{\tau}
\def\up{\upsilon}
\def\ph{\phi}
\def\vp{\varphi}
\def\ch{\chi}
\def\ps{\psi}
\def\om{\omega}
\def\Ga{\Gamma}
\def\De{\Delta}
\def\Th{\Theta}
\def\La{\Lambda}
\def\Si{\Sigma}
\def\Up{\Upsilon}
\def\Ph{\Phi}
\def\Ps{\Psi}
\def\Om{\Omega}
\def\mn{{\mu\nu}}
\def\cD{{\cal D}}
\def\cF{{\cal F}}
\def\cL{{\cal L}}
\def\cS{{\cal S}}
\def\fr#1#2{{{#1} \over {#2}}}
\def\frac#1#2{\textstyle{{{#1} \over {#2}}}}
\def\pt#1{\phantom{#1}}
\def\prt{\partial}
\def\vev#1{\langle {#1}\rangle}
\def\ket#1{|{#1}\rangle}
\def\bra#1{\langle{#1}|}
\def\amp#1#2{\langle {#1}|{#2} \rangle}
\def\half{{\textstyle{1\over 2}}}
\def\lsim{\mathrel{\rlap{\lower4pt\hbox{\hskip1pt$\sim$}}
    \raise1pt\hbox{$<$}}}
\def\gsim{\mathrel{\rlap{\lower4pt\hbox{\hskip1pt$\sim$}}
    \raise1pt\hbox{$>$}}}
\def\ol#1{\overline{#1}}
\def\Re{\hbox{Re}\,}
\def\Im{\hbox{Im}\,}
\def\etal {{\it et al.}}
\def\slash#1{\not\hbox{\hskip -2pt}{#1}}

\section{Introduction}
The observation that the Universe appears to be accelerating has been one
of the biggest surprises in both the astronomy and particle physics 
communities.
This acceleration can be accounted for by adding
a cosmological constant in Einstein's
equations. While particle physicists never had a definite reason
for setting the cosmological constant to zero, there was always the view that
like all small or zero quantities, there must be a symmetry enforcing it.
The observation that the size of the cosmological constant is neither zero
nor at one of the natural scales like the Planck length or the scale of 
supersymmetry breaking has focused attention on explaining the dark energy 
component. 

There are at least two requirements for any model of dark energy: (1) one must 
obtain in a natural way the observed size of the energy density, 
$\rho \sim 3\times 10^{-3}$~eV$^4$. Many models attempt to relate this to 
the observed coincidence of this scale with the scale $H_0^2 M_{Pl}^2$ where
$H_0$ is the Hubble constant at the current epoch and $M_{Pl}$ is the 
Planck mass. (2) One must also obtain an equation of state $w$ which is 
similar to that of a cosmological constant at least in the recent past, 
namely $w\approx -1$. In fact, the most
recent data from the Wilkinson Microwave Anisotropy Probe
(WMAP) satellite and supernova and sky surveys 
have constrained the equation of state 
$-1.4 < w < -0.8$ at the 95\% confidence level for dark 
energy with a constant $w$. 

An interesting approach to the dark matter problem that has arisen from recent
advances in understanding string theory and black hole physics is to employ 
a holographic principle. Here the motivation is the recognition that the 
observed size of the energy density at the present epoch seems to be 
consistent with taking the geometric average of the two scales $H_0$ and 
$M_{Pl}$. Early attempts to relate the dark energy to the
Hubble scale or the particle horizon did not give acceptable solutions in 
detail that were consistent with the observations.

The evolution of the dark energy density also depends on its precise nature.
If the dark energy is like a cosmological constant with equation of state 
$w_\Lambda =-1$, the dark energy density is constant and within the expanding 
Universe ultimately the dark energy density comes to dominate over the matter
density. In this scenario the transition from matter domination to dark energy
domination is rapid, and the fact that experimental observations seem to be
consistent with sizeable amounts of both matter and dark energy indicates that
we are currently residing in this transition period. The cosmic coincidence 
problem is the statement that it is unlikely that the current epoch coincides
with this rapid transition period.

A possible solution is to assume that there is some mechanism that is 
converting dark energy into matter. 
Cosmic antifriction forces were introduced in Ref.~\cite{Zimdahl:2000zm} as 
a possible solution to the cosmic coincidence problem. If these forces are 
present one can obtain a fixed ratio of matter density to dark energy density
as the final state of the Universe and hence solve the coincidence problem.
The 
resulting cosmology is described by Friedmann equations.
The basic scenario is quite simple: the natural tendency of a cosmological 
component with a more negative equation of state to dominate at large times 
the energy density of the Universe is compensated by the decay of the 
dark energy 
component into the other component(s). At large times an equilibrium 
solution can develop for which the various components can coexist. The
interpretation is then that we are living in or close to this equilibrium thus
eliminating the cosmic coincidence problem.

Frictional and antifrictional forces we envision here were introduced for 
a single component in Refs.~\cite{Zimdahl:2000zm,Balakin:2003tk}. The 
particular case where the dark energy component is assumed to have a constant
equation of state was treated in Ref.~\cite{Zimdahl:2005bk}.
These efforts were preceded by searches for attractor solutions involving 
scalar fields\cite{Amendola:1999qq,Amendola:2000uh}. Recently these kind of 
scenarios have been employed in models with brane-bulk energy 
exchange\cite{Elizalde:2005ju}.
Some attempts have been made
to explain the size of the dark energy density on the basis
of holographic ideas\cite{Cohen:1998zx}, and lead to the consideration of 
the Hubble horizon as the holographic scale\cite{Horava:2000tb,Thomas:2002pq}. 
A holographic model based on the future event horizon was examined in 
Ref.~\cite{Li:2004rb}. Danielsson studied the effects of a transplanckian
backreaction as a particular source for the holographic dark 
energy\cite{Danielsson:2004xw}. More recenlty 
an emphasis on the case of interacting dark energy has been discussed 
in Refs.~\cite{Myung:2005pw,Kim:2005at,Kim:2005gk}.

Ultimately one needs to appeal to the observational data to constrain these
ideas. In this direction 
Wang, et al. have begun a more detailed comparison of the predictions
of a holographic universe with interaction with the observational 
data\cite{Wang:2005ph}. In addition to the overall development of the 
densities of the cosmological components there are other observables which 
are sensitive to the modified scenarios\cite{Zimdahl:2005bk,Wang:2004cp}.

\section{Framework of Interacting Dark Energy}

We assume two component equations for dark energy and matter,
\begin{eqnarray}
&&\dot{\rho}_\Lambda + 3H(1+w_\Lambda)\rho_\Lambda = -Q\;, \nonumber \\
&&\dot{\rho}_m + 3H(1+w_m)\rho_m = Q\; .
\label{twocomp}
\end{eqnarray}
The equality of $Q$ in the two equations guarantees the overall
conservation of the energy-momentum tensor, and positive $Q$ can be 
interpreted as a transfer from the dark energy component to the matter 
component. Presumably this arises from some microscopic mechanism, but we do 
not specify one here. The quantity $\rho_m$ is the matter component for which
one usually takes $w_m=0$, but we temporarily retain a
nonzero equation of state so as to retain the generality of an two interacting
components.

One can define effective equations of state,
\begin{eqnarray}
&&\dot{\rho}_\Lambda + 3H(1+w_\Lambda^{\rm eff})\rho_\Lambda = 0\;, 
\nonumber \\
&&\dot{\rho}_m + 3H(1+w_m^{\rm eff})\rho_m = 0\;
\label{definew}
\end{eqnarray}
Define the ratio $r=\rho_m/\rho_\Lambda$ and the rate
$\Gamma =Q/\rho_\Lambda$, then the effective equations of state are expressed
in terms of the native equations of state and a ratio of rates as
\begin{eqnarray}
w_\Lambda ^{\rm eff}=w_\Lambda+{{\Gamma}\over {3H}}\;, \qquad
w_m ^{\rm eff}=w_m-{1\over r}{{\Gamma}\over {3H}}\;.
\end{eqnarray}
The time evolution of the ratio $r$ is then
\begin{eqnarray}
\dot{r}=3Hr\left [w_\Lambda-w_m+{{1+r}\over r}{{\Gamma}\over {3H}}\right ]
=3Hr\left [w_\Lambda^{\rm eff}-w_m^{\rm eff}\right ].
\label{rdot}
\end{eqnarray}
The general behavior of solutions can be ascertained from this equation. 
The stable points in the evolution are obtained when either $r=0$ or when
$w_\Lambda^{\rm eff}=w_m^{\rm eff}$. If $w_\Lambda^{\rm eff}< w_m^{\rm eff}$
then the vanishing dark energy density ($r=0$) solution will apply in the 
infinite past while a constant nonzero ratio of dark energy to matter will 
apply in the infinite future with equal effective equations of state.

Using the definitions
\begin{eqnarray}
\Omega_\Lambda={{8\pi\rho_\Lambda}\over {3M_p^2H^2}}\;, \qquad
\Omega_m={{8\pi\rho_m}\over {3M_p^2H^2}}\;, 
\end{eqnarray}
which satisfy a Friedmann equation 
\begin{eqnarray}
\Omega_\Lambda +\Omega_m=1\;,
\end{eqnarray}
one can convert to the physical parameters
\begin{eqnarray}
r={{1-\Omega_\Lambda}\over {\Omega_\Lambda}}\;, \qquad 
\dot{r}=-{{\dot{\Omega}_\Lambda}\over {\Omega_\Lambda^2}}\;.
\label{randomega}
\end{eqnarray}

One can write the differential 
equation in the suggestive form
using $\dot{\Omega}_\Lambda=H{{d\Omega_\Lambda}\over {dx}}$,
\begin{eqnarray}
{{d\Omega_\Lambda}\over {dx}}=-3\Omega_\Lambda(1-\Omega_\Lambda)
\left [w_\Lambda^{\rm eff}-w_m^{\rm eff}\right ]\;.
\label{diffom}
\end{eqnarray}
where $x=\ln (a/a_0)$.
The utility of the evolution equation in this form is that it allows one to
understand in a general way the possible solutions for $\Omega _\Lambda$. 
Firstly if one turns off the interaction ($\Gamma =0$) one recovers the 
usual evolution equation with the effective equations of state replaced with
the usual equations of state $w_\Lambda$ and $w_m$ (we will call these 
the native equations of state). For constant $w_\Lambda$ and $w_m$ and 
$w_\Lambda < w_m$ then the dark energy evolution proceeds from the fixed 
points of the differential equation (zeros of the right hand side) in the usual
way from $\Omega_\Lambda=0$ to $\Omega_\Lambda =1$ in the usual way. When 
an interaction is present a more interesting solution may occur where the 
evolution of $\Omega_\Lambda$ approaches a condition where instead
$w_\Lambda^{\rm eff}=w_m^{\rm eff}$ at which $\Omega _\Lambda $ approaches 
a fixed asymptotic value for large time less than 1. For a properly chosen
interaction this might constitue a solution to the cosmic concidence problem.
 
For illustration one can also consider the (redundant) differential equation 
for the matter density,
\begin{eqnarray}
{{d\Omega_\Lambda}\over {dx}}=-3\Omega_\Lambda \Omega_m
\left [w_\Lambda^{\rm eff}-w_m^{\rm eff}\right ]\;, \nonumber 
\\      
{{d\Omega_m}\over {dx}}=-3\Omega_m \Omega_\Lambda
\left [w_m^{\rm eff}-w_\Lambda^{\rm eff}\right ]\;.
\end{eqnarray}
An initial condition $(\Omega_m,\Omega_\Lambda)=(1,0)$ is inevitably 
driven to an asymptotic solution for which $w_m^{\rm eff}=w_\Lambda^{\rm eff}$
and $\Omega_m$ and $\Omega_\Lambda$ approach their limiting (equal) value. 
This 
behavior is displayed for a specific solution\cite{Kim:2005at} shown in Fig.~1.

\bigskip
\begin{figure}[h]
\centerline{
\mbox{\includegraphics[width=3.50in]{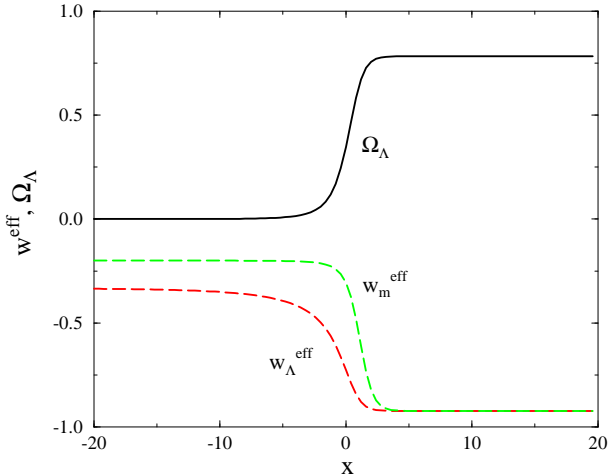}}
}
\caption{Evolution of $\Omega_\Lambda$ (solid)
and the effective equations of state,
$w_\Lambda^{\rm eff}$ (dashed, bottom) and $w_m^{\rm eff}$ (dashed, top). 
The horizontal 
axis is $x=\ln (a/a_0)$ for an arbitrary $a_0$. The future event horizon (FH)
is assumed to be the length scale for determining $w_\Lambda^{\rm eff}$, while
the interaction is given by $\Gamma/3H=b^2/\Omega$ with $b^2=0.2$. }
\label{overView}
\end{figure}
\bigskip

Two physical assumptions must be provided to determine the evolution of 
the parameters of the universe. A holographic condition on the dark energy
$\rho_\Lambda$  will determine by itself the effective equation of state
$w_\Lambda ^{\rm eff}$. A specific specification for the source $Q$ will 
determine the effective matter equation of state $w_m^{\rm eff}$.
Alternatively one of these conditions can be replaced by the assumption that
$w_\Lambda$ is constant.

One typically obtains an equation for the evolution of the dark energy density
of the form 
\begin{eqnarray}
{{d\Omega_\Lambda}\over {dx}}=f(\Omega_\Lambda)\;.
\label{diffeq}
\end{eqnarray}
The evolution of $\Omega_\Lambda $ with respect to $a$ or $x$ is qualitatively
determined by the fixed points of the differential equation in 
Eq.~(\ref{diffeq}) which are determined by the condition $f(\Omega_\Lambda)=0$.
Examining the equations in Eqs.~(\ref{randomega}), it would appear that this 
condition would require that $\dot{r}=0$ and then that 
$w_\Lambda^{\rm eff}=w_m^{\rm eff}$ via Eq.~\ref{rdot}.
However, as shown in Ref.~\cite{Kim:2005at} if one of the fixed points occurs
for the physically interesting value of $\Omega_\Lambda=0$, then one 
can have the Universe with close to vanishing dark energy simultaneously
with $w_\Lambda^{\rm eff}\ne w_m^{\rm eff}$.
In the remainder of this paper we explore the general conditions that give
rise to this type of equilibrium solution.

\section{Dimensional analysis}

First define a length scale $L_\Lambda$ via
\begin{eqnarray}
\rho_\Lambda={{3c^2M_p^2}\over {8\pi L_\Lambda^2}}
\label{definel}
\end{eqnarray}
where the constant $c$ represents an order one constant\cite{footnote1}.
Specific assumptions
might associate the 
length scale with a physical length such as the Hubble horizon, the particle
horizon, the future event horizon, or perhaps some other parameter.
The common feature of any realistic choice is that the length scale at the 
present epoch should be of order the size of the 
observable Universe so as to obtain 
the appropriate size of the dark energy density consistent with experimental 
observations. One has
\begin{eqnarray}
L_\Lambda={c\over {H\sqrt{\Omega_\Lambda}}}\;.
\label{llambda}
\end{eqnarray}

There need to be two constraints imposed to solve the set of equations. We 
have three choices: 1) holographic principle on $\rho_\Lambda$, 
2) suppose a form 
for $Q$ or equivalently $\Gamma$, and 3) suppose a form for the equation of 
state $w_\Lambda$. The equation linking them is obtained from 
Eq.~(\ref{twocomp}),
\begin{eqnarray} 
\Gamma =3H(-1-w_\Lambda)
+2{{\dot{L}_\Lambda}\over {L_\Lambda}}
\label{rates}
\end{eqnarray}
This implies the generic scale of the decay rate $\Gamma$ is the same order 
as $H$. In the particular case $w_\Lambda=-1$ Eq.~(\ref{rates}) reduces to 
\begin{eqnarray}
\Gamma =2{{\dot{L}_\Lambda}\over {L_\Lambda}}
\end{eqnarray}
which connects the sign of $Q$ to the sign of 
${{\dot{L}_\Lambda}\over {L_\Lambda}}$.
In the case that the interaction vanishes, $\Gamma=0$, one returns to the 
condition of a cosmological constant with constant energy density (and 
therefore constant $L_\Lambda$).
On the other hand the interaction can either increase or decrease 
$\rho_\Lambda$ depending on the sign of the interaction.
Henceforth we will refer to the rate ${{\dot{L}_\Lambda}\over {L_\Lambda}}$
as the holographic rate, but it should be understood that it is an equivalent
description for the time development of the dark energy density independent of 
any assumption based on the holographic principle.

Since the dark energy density scales as 
$\rho_\Lambda \propto {1\over L_\Lambda^2}$,
one has
\begin{eqnarray}
{\dot{\rho}_\Lambda\over \rho_\Lambda}=-2{{\dot{L}_\Lambda}\over {L_\Lambda}}
\;.
\end{eqnarray}
Comparing to Eq.~(\ref{definew}) one obtains the relationship between the
effective equation of state and the length scale
\begin{eqnarray}
w_\Lambda^{\rm eff}=-1+{2\over 3H}{{\dot{L}_\Lambda}\over {L_\Lambda}}\;.
\label{general}
\end{eqnarray}
This equation recasts the scaling exhibited by the dark energy
in terms of 
the ratio of the two rates, $\dot{L}_\Lambda/L_\Lambda$ and $H$. When 
there is no 
interaction the same equation applies for the native equation of state 
$w_\Lambda$. The 
familiar case of a constant dark energy density simply corresponds to the 
statement that the length scale does not change, $\dot{L}_\Lambda/L_\Lambda=0$.
More generally Eq.~(\ref{general}) shows that an assumption for $L_\Lambda$ 
based on some holographic principle determines the form of the effective 
equation of state even in the presence of an interaction.
If one applies Eq.~(\ref{llambda}) one has 
\begin{eqnarray}
{{\dot{L}_\Lambda}\over {L_\Lambda}}=-{\dot{H}\over H}-{1\over 2}
{{\dot{\Omega}_\Lambda}\over {\Omega_\Lambda}}
\label{logderiv}
\end{eqnarray}
Inserting this into Eq.~(\ref{general}) and using the definition for the 
deceleration parameter $q=-a\ddot{a}/\dot{a}^2$ and $\dot{H}/H^2=-1-q$, 
one obtains 
\begin{eqnarray}
w_\Lambda^{\rm eff}=-{1\over 3}+{2\over 3}q-{1\over 3H}
{{\dot{\Omega}_\Lambda}\over {\Omega_\Lambda}}
\end{eqnarray}
Asymptotically (i.e. for large $x$) the last term vanishes, so that 
\begin{eqnarray}
w_\Lambda^{\rm eff}\:\: \rightarrow \:\: -{1\over 3}+{2\over 3}q=
-1-{2\over 3}{\dot{H}\over H^2}\;.
\label{asymp}
\end{eqnarray}

Finally using Eq.~(\ref{logderiv}) one obtains
\begin{eqnarray}
{{d\Omega_\Lambda}\over {dx}}=-2\Omega_\Lambda\left [ {\dot{H}\over H^2}
+{1\over H}{{\dot{L}_\Lambda}\over {L_\Lambda}}\right ]
\label{diffph}
\end{eqnarray}
which indicates that the differential equation (right hand side) has 
a zero for $\Omega_\Lambda =0$ provided the term in brackets 
does not vanish like $1/\Omega_\Lambda$.

The resulting interpretation of the possible evolutions of the Universe becomes
straightforward using Eq.~(\ref{diffph}). 
The derivative of $\Omega_\Lambda$ has
fixed points at $\Omega_\Lambda=0$ and at some larger value where the 
condition 
\begin{eqnarray}
{\dot{H}\over H^2}
+{1\over H}{{\dot{L}_\Lambda}\over {L_\Lambda}}=0\;,
\end{eqnarray}
is satisfied. For large times the Universe approaches this solution where the
effective equations of states for matter and the dark energy become equal. At 
early times the dark energy is repelled from the $\Omega_\Lambda=0$ and grows 
monitonically to a positive value. For the case $\dot{L}_\Lambda/{L_\Lambda}>0$
one observes that $\dot{H}/H^2=-1-q$ must be negative for asympotically
large times.

\section{Examples}

A holographic condition based on relating $L_\Lambda$ to the Hubble scale or 
to the particle horizon leads to conflict with observation. In the former case
the ratio of the dark energy density to matter density is constant. In the 
latter case the equation of state of the dark energy is greater than $-1/3$. 
Taking the length scale to be the Hubble 
horizon $L_\Lambda=R_{HH}=1/H$ which implies 
\begin{eqnarray}
\rho_\Lambda={{3c^2M_p^2H^2}\over {8\pi}}\;,
\end{eqnarray}
so that $c^2=\Omega_\Lambda$ is a constant.
This yields
\begin{eqnarray}
w_\Lambda ^{\rm eff}=-{1\over 3}+{2\over 3}q\;,
\end{eqnarray}
so that the asymptotic solution obtained in Eq.~(\ref{asymp}) is identically 
satisfied.

Setting the holographic length scale equal to the particle horizon gives
\begin{eqnarray}
R_{\rm PH}=a\int_0^t{dt\over a}=a\int_0^a {da\over Ha^2}\;.
\end{eqnarray}
while setting it equal to the future event 
horizon\cite{Li:2004rb} yields
\begin{eqnarray}
R_{\rm FH}=a\int_t^\infty {dt\over a}=a\int_a^\infty {da\over Ha^2}\;.
\end{eqnarray}

If the length scale is set equal to the Hubble horizon, the particle 
horizon, or the future horizon, then the rates go like
\begin{eqnarray}
{1\over H}{{\dot{L}_\Lambda}\over {L_\Lambda}} 
&=&-{\dot{H}\over H^2}\quad {\rm (HH)}\;, \nonumber \\
&=&1+{\sqrt{\Omega_\Lambda}\over c}\quad {\rm (PH)}\;, \nonumber \\
&=&1-{\sqrt{\Omega_\Lambda}\over c}\quad {\rm (FH)}\;,
\label{holocase}
\end{eqnarray}
respectively.
Finally for a cosmological constant one has 
${{\dot{L}_\Lambda}/{L_\Lambda}}=0$.
These conditions yield the following expressions for the 
effective dark energy equation 
of state
\begin{eqnarray}
w_\Lambda ^{\rm eff}&=&-1-{2\over 3}{\dot{H}\over H^2}
\quad {\rm (HH)}\;, \nonumber \\
&=& -{1\over 3}+{2\over 3}
{\sqrt{\Omega_\Lambda}\over c}\quad {\rm (PH)}\;, \nonumber \\
&=&-{1\over 3}-{2\over 3}{\sqrt{\Omega_\Lambda}\over c}\quad {\rm (FH)}\;,
\label{effectivecase}
\end{eqnarray}
and $w_\Lambda ^{\rm eff} =-1$ for a cosmological constant.

The case of the Hubble horizon is particularly simple: From 
Eq.~(\ref{logderiv}) one has that $\dot{\Omega}_\Lambda=0$, so that the ratio
of dark energy to matter is constant\cite{Hsu:2004ri}. This is independent of 
the existence of a interaction $\Gamma$. If the interaction is absent, then
$w_\Lambda^{\rm eff}=0$ and $\dot{H}/H^2=-3/2$ identically. Including an 
interaction these generalize to 
\begin{eqnarray}
w_\Lambda^{\rm eff}=w_m^{\rm eff}=-{\Gamma\over {3H}}
{{\Omega_\Lambda}\over {1-\Omega_\Lambda}}=-{{\Gamma}\over {3H}}{1\over r}
\end{eqnarray}

One particular mechanism for determining the holographic condition is to 
assume the dark energy density is created out of the vacuum as the Universe 
expands\cite{Danielsson:2004xw}. The component is created via the Bogolubov
modes which result from the expansion of the Universe and the variation of the 
vacuum in such a situation. One has
\begin{eqnarray}
\rho_\Lambda={3\over {2\pi}}{{\Lambda^2}\over {a^4}}
\int_{a_i}^a \: dx\: x^3H^2(x)\;,
\label{bog}
\end{eqnarray}
where $\Lambda$ is some mass of order the Planck scale.
Translating this into the varying scale $L_\Lambda$, one obtains
\begin{eqnarray}
L_\Lambda=a^2\left [\int_{a_i}^a \: dx\: x^3H^2(x)\right ]^{-1/2}\;,
\end{eqnarray}
where an overall constant that arises in this definition is absorbed into the 
definition of the order one constant coefficient $c$. One calculates 
\begin{eqnarray}
{1\over H}{{\dot{L}_\Lambda}\over {L_\Lambda}} 
=2-{1\over 2}H^2L_\Lambda^2=2-{1\over 2}{c^2\over \Omega_\Lambda} \;.
\end{eqnarray}
The effective equation of state then follows from Eq.~(\ref{general}). 
In this formulation the behavior near $\Omega_\Lambda=0$ results from 
lower momentum modes which were created at earlier times when $H$ was larger.

The asymptotic value for the effective equation of state should approach 
something near that of a cosmological constant to be consistent with 
oberservations. Having accumulated the results for the previous examples it
is easy to conclude that the future horizon\cite{Li:2004rb}
yields an acceptable effective equation of state near $-1$, while the particle
horizon has an effective equation of state which is too large.
Generally one wants a length scale rate change that becomes small compared to 
the Hubble expansion $\dot{L}_\Lambda/L_\Lambda < H$ so that the effective
equation of state approaches $-1$ as in Eq.~(\ref{general}). The future horizon
and other holographic conditions that result in a decreasing value for 
$\dot{L}_\Lambda /L_\Lambda$ as a function of increasing $\Omega_\Lambda$ 
have the desired properties for an accelerating Universe.

\section{Interactions}

The coupled equations 
in Eq.~(\ref{twocomp}) defines an interaction between the dark 
energy and matter components.
Given a specific rate $\Gamma$ as a function of $H$ and $\Omega_\Lambda$, 
one can study the behavior
of the evolution equations with a specific behavior on the energy density.
The ratio of rates is taken to be a function of $\Omega _\Lambda$
\begin{eqnarray}
{\Gamma\over {3H}}=b^2g(\Omega_\Lambda)\;,
\label{gdef}
\end{eqnarray}
where the dimensionless constant $b^2$ is included to facilitate comparison
with previous works. In particular Ref.~\cite{Kim:2005at} assumed an
interaction of this form with $g(\Omega_\Lambda)=1/\Omega_\Lambda$.
One then obtains (we assume henceforth that $w_m=0$) 
\begin{eqnarray}
w_m^{\rm eff}=-{\Gamma \over {3H}}{{\Omega_\Lambda}\over 
{(1-\Omega_\Lambda)}}\;.
\label{generalm}
\end{eqnarray}
So the effective equation of state $w_m^{\rm eff}$
for the matter component depends on the 
interaction only. Together with the fact that the effective equation of state
$w_\Lambda^{\rm eff}$ depends only on the assumption for the physical 
condition setting the scale $L_\Lambda$, allows for a more general analysis 
of possible cases that give the desired equilibrium solution.
Using Eqs.~(\ref{rdot}) and (\ref{general}) one obtains
\begin{eqnarray}
\dot{r}=3Hr\left [-1+{2\over 3H}{{\dot{L}_\Lambda}\over {L_\Lambda}}
+{\Gamma \over {3H}}{{\Omega_\Lambda}\over 
{(1-\Omega_\Lambda)}}\right ]\;.
\end{eqnarray}
Using Eq.~(\ref{randomega}) to convert this into an equation involving 
only $\Omega_\Lambda$, one obtains
\begin{eqnarray}
{{d\Omega_\Lambda}\over {dx}}=3\Omega_\Lambda(1-\Omega_\Lambda)
\left [1-{2\over 3H}{{\dot{L}_\Lambda}\over {L_\Lambda}}
-{\Gamma \over {3H}}{{\Omega_\Lambda}\over 
{(1-\Omega_\Lambda)}}\right ]
\label{diffgen}
\end{eqnarray}
This expression is quite general, so it facilitates a more general 
understanding of the physical conditions required for the holographic 
conditions determining $L_\Lambda$ and the interaction $\Gamma$.
If these conditions have a functional expression in terms of 
$\Omega _\Lambda$, then Eq.~(\ref{diffgen}) represents a differential 
equation that can be solved. This is the case for the 
holographic conditions arising from the particle and future horizons as well
as the form
of the interactions assumed in this paper.
Furthermore the differential equation involves the effective equations of 
state in a simple way. 
The first two terms in brackets are $-w_\Lambda^{\rm eff}$
while the last term is $w_m^{\rm eff}$.
Together with Eq.~(\ref{diffph}) it determines the 
acceleration of the Universe through $\dot{H}/H^2=-1-q$.
One has\footnote{Note that the quantity $\dot{H}/H$ is not another independent 
rate, but is determined in terms of the fundamental rates $H$, 
$\dot{L}_\Lambda /L_\Lambda$, and $\Gamma$.}
\begin{eqnarray}
{\dot{H}\over {H^2}}=-{3\over 2}(1-\Omega_\Lambda)-{\Omega_\Lambda \over H}
{{\dot{L}_\Lambda}\over {L_\Lambda}}+{\Gamma \over {2H}}
\Omega_\Lambda\;.
\label{accelgen}
\end{eqnarray}

A complete specification of the problems involves the specification of 
two ratios of rates as in Eq.~(\ref{rates}). A generic choice of definition 
for $\dot{L}_\Lambda/L_\Lambda$ and $\Gamma$ will determine the native 
equation of state $w_\Lambda$ which will then vary with time. Alternatively
one can choose a constant $w_\Lambda$ and $\dot{L}_\Lambda/L_\Lambda$ in which
case the interaction rate is determined. We examine these cases in more detail
in the remainder of this section.

\subsection{Constant equation of state}

The first case we shall examine is the one where the native equation of state
for dark energy is constant. For this purpose take the equation involving the 
three rates, Eq.~(\ref{rates}), and eliminate the interaction term in 
Eq.~(\ref{diffgen}) in favor of the other two rates. One obtains
\begin{eqnarray}
{{d\Omega_\Lambda}\over {dx}}=3\Omega_\Lambda
\left [1+w_\Lambda \Omega_\Lambda
-{2\over 3H}{{\dot{L}_\Lambda}\over {L_\Lambda}}
\right ]
\label{diffgenw}
\end{eqnarray}
This equation is valid irregardless of whether $w_\Lambda$ is constant in time,
but we are presently interested in the constant case.
Provided the holographic rate does not have a singularity for $\Omega_\Lambda$,
the right hand side has the required zero at $\Omega_\Lambda=0$, and
for $0<\Omega_\Lambda <1$ the first two terms in brackets is positive provided 
$w_\Lambda \Omega_\Lambda > -1$. 
For both the case of the holographic rate based on the particle horizon and 
future event horizon, there is another fixed point at positive 
$\Omega_\Lambda$ (see Eq.~(\ref{holocase})).

Using Eqs.~(\ref{general}) and (\ref{diffgenw}) one obtains the simple 
result
\begin{eqnarray}
{{d\Omega_\Lambda}\over {dx}}=3\Omega_\Lambda
\left [w_\Lambda \Omega_\Lambda
-w_\Lambda ^{\rm eff}
\right ]\;,
\label{diffgenw2}
\end{eqnarray}
which shows that the fixed point occurs when 
$w_\Lambda ^{\rm eff}=w_\Lambda \Omega_\Lambda$.
As examples take $w_\Lambda=-1$ and $c=1$. Then the attractive fixed point
occurs at $\Omega_\Lambda ={1\over 9}$ for the particle horizon (PH) 
case and at $\Omega _\Lambda =1$ for the future event horizon (FH) case, 
and from Eq.~(\ref{effectivecase}) these correspond to 
$w_\Lambda ^{\rm eff}=-{1\over 9}$
and $-1$ respectively. More generally Eq.~(\ref{diffgenw2}) implies that 
$w_\Lambda ^{\rm eff}\le w_\Lambda$.

In the case the scale $L_\Lambda $ is identified with the Hubble scale $1/H$, 
the differential equation vanishes and one has the relation
\begin{eqnarray}
w_\Lambda ^{\rm eff}=w_\Lambda \Omega_\Lambda \;.
\end{eqnarray} 
If one further assumes a 
constant $w_\Lambda$ then the interaction $\Gamma$ is determined, and 
$w_\Lambda^{\rm eff}$ is simply proportional to the constant $\Omega_\Lambda$.
The interaction rate is then proportional to $\Omega_m$ as shown in 
Ref.~\cite{Pavon:2005yx},
\begin{eqnarray}
{\Gamma\over {3H}}=-w_\Lambda(1-\Omega_\Lambda)\;.
\end{eqnarray}
So this holographic condition corresponds to the one in which the differential
equation is identically zero and equilibrium between the dark energy density 
and matter density holds for all times.

\subsection{Assumptions about the holographic and interaction rates}

The inclusion of an iteraction term for the dark energy component can be 
accomodated most easily by defining an effective equation of state
$w_\Lambda^{\rm eff}$ which differs from the native equation of state
$w_\Lambda$. By virtue of its definition it satisfies the same equation that 
the native equation of state satisfies in the noninteraction case, namely
Eq.~(\ref{general}).
If $0 < \dot{L}_\Lambda/L_\Lambda < H$ 
then $-1 < w_\Lambda^{\rm eff} < -1/3$ and
the universe accelerates. The effective equation of state must lie between
that of a cosmological constant ($w=-1$) and the curvature component
($w=-1/3$).

While there is a certain appeal in taking the future event horizon as the 
length scale for a holographic prinicple\cite{Li:2004rb}, 
it is clear from the evolution 
equation for the dark energy that an equilibrium fixed point solution can 
occur for a wider variety of assumptions. 

In order to examine the conditions on the interaction rate consider the cases
\begin{eqnarray}
{\Gamma \over {3H}}={{b^2}\over {\Omega_\Lambda^n}}\;,
\label{int}
\end{eqnarray}
for some exponent $n$.
One then obtains 
\begin{eqnarray}
w_m^{\rm eff}=-{{b^2}\over {(1-\Omega_\Lambda)\Omega_\Lambda^{n-1}}}\;.
\end{eqnarray}
Figure 2 shows the functional dependence of $w_\Lambda^{\rm eff}$ and 
$w_m^{\rm eff}$ versus $\Omega_\Lambda$ for various choices of the holographic
condition and interaction. An intial condition of $\Omega _\Lambda=0$ 
will evolve to the right until $w_\Lambda^{\rm eff}=w_m^{\rm eff}$ which 
respresents the asymptotic solution.
It is clear that an interaction of the form 
$\Gamma /3H=b^2/\Omega_\Lambda^n$ with $n\leq 1$ can give an acceptable
result with a equilibrium balance between $\Omega_\Lambda$ and 
$\Omega_m=1-\Omega_\Lambda$ provided the holographic condition involves the
future event horizon. 

\bigskip
\begin{figure}[h]
\centerline{
\mbox{\includegraphics[width=3.50in]{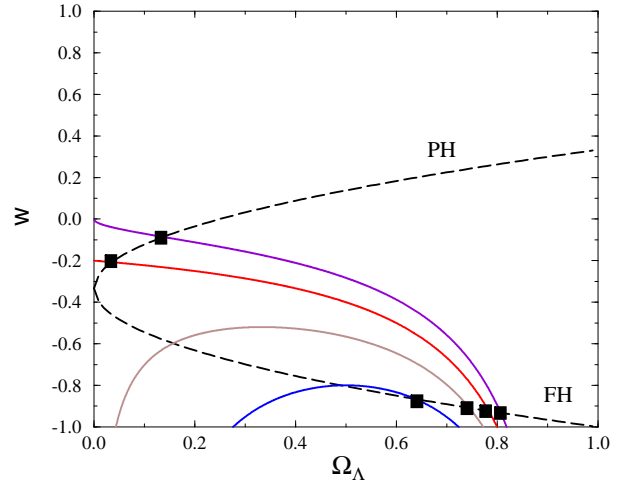}}
}
\caption{Dependence of $w_\Lambda^{\rm eff}$ (dashed)
for the case of particle horizon (PH, top) and future event horizon
(FH, bottom) and $w_m^{\rm eff}$ (solid) for 
$g(\Omega_\Lambda)=1/\Omega_\Lambda ^n$ for $n={1\over 2},1,{3\over 2},2$ 
from top to bottom. The fixed points for the differential equation occur 
when $w_\Lambda^{\rm eff} =w_m^{\rm eff}$, and the attractive ones for 
large $x$ are indicated with filled squares. For the initial condition 
$\Omega_\Lambda=0$, the choice of FH and $n\le 1$ gives an evolution that could
be consistent with observations. The parameters were
assigned as $c=1$ and $b^2=0.2$.} 
\label{evolution}
\end{figure}
\bigskip

For $n>1$ the effective equation of state $w_m^{\rm eff}$ diverges for 
small $\Omega_\Lambda$, and in fact is less than $w_\Lambda ^{\rm eff}$ so 
does not yield an acceptable solution.
We note that for $n<1$ one has $w_m^{\rm eff}=0$ for $\Omega_\Lambda=0$ which 
may be needed for better agreement with the observational data.
An explicit example is plotted in Fig.~3 for which the iteraction is assumed to
be of the form in Eq.~(\ref{int}) with $n=1/2$. In fact this is the typical 
behavior for all $n<1$.

\bigskip
\begin{figure}[h]
\centerline{
\mbox{\includegraphics[width=3.50in]{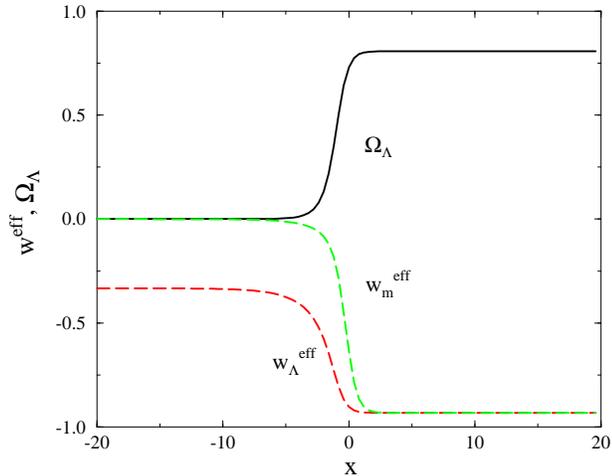}}
}
\caption{Evolution of $\Omega_\Lambda$ (solid)
and the effective equations of state,
$w_\Lambda^{\rm eff}$ (dashed, bottom) and $w_m^{\rm eff}$ (dashed, top) for 
an interaction given by $n=1/2$. 
The horizontal 
axis is $x=\ln (a/a_0)$ for an arbitrary $a_0$. The parameters used are the 
same as given in Figure 2.}
\label{overView}
\end{figure}
\bigskip

\section{Multicomponent Generalization}

An obvious generalization of the model considered here is the multicomponent 
case 
\begin{eqnarray}
&&\dot{\rho}_\Lambda + 3H(1+w_\Lambda)\rho_\Lambda = -Q\;, \nonumber \\
&&\dot{\rho}_i + 3H(1+w_i)\rho_i = Q_i\;,
\label{multicomp}
\end{eqnarray}
subject to the constraint $\sum_i Q_i=Q$. The evolution equation for the 
dark energy component is
\begin{eqnarray}
{{d\Omega_\Lambda}\over {dx}}=-3\Omega_\Lambda \sum_i \Omega_i
\left [w_\Lambda^{\rm eff}-w_i^{\rm eff}\right ]\;
\end{eqnarray}
where 
\begin{eqnarray}
w_i^{\rm eff}=w_i-{1\over r_i}{\Gamma _i\over 3H}, \quad 
r_i={\rho_i\over \rho_\Lambda}, \quad
 \sum_i\Omega _i=1-\Omega_\Lambda,
\end{eqnarray}
are quantities defined for each component other than the dark energy.
The generic solution that occurs for typical cases where the effective 
equations of state are monotonic functions of $\Omega_\Lambda$ is that
two of the components come to equilibrium with
the remaining components being diluted away by the expansion of the 
Univerese. This case is of coursed realized if one adds a radiation component
for example.

\section{Summary}

This work presents a unifying and general treatment of the physical 
assumptions that have been made for models with 
an interacting dark energy component in the 
Universe. The discussion in the literature has typically 
involved specifications  
for a holographic principle, a choice for an interaction between
dark matter and dark energy, or the assumption of 
constant equations of state. A specification
for any two of the three choices determines the third. 
In particular a generic specification of a holographic principle and 
an interaction gives an effective equation of state that is not constant.
The existence of a future fixed point for the 
dark energy of the Universe can be easily established without a numerical 
calculation by analyzing Eq.~(\ref{diffgen}) with appropriate assumptions 
for the relative size of the rate for the Universe expansion $H$, 
the rate for the evolution of the dark energy density 
$\dot{L}_\Lambda/L_\Lambda$, and the interaction rate $\Gamma$.
Furthermore if the Universe in the present epoch is at or near this fixed 
point, a condition on the parameters involved in these physical assumptions
can be derived by the vanishing of this differential equation.
The evolution of a Universe with vanishing dark energy can be understood 
in a qualitative fashion if the interaction rate and the holographic rate
or plotted versus as shown in Fig.~2, and the asymptotic future state of 
the Universe is identified as the balance between two competing factors: 
the natural tendency for dark energy to dominate over matter as the Universe
expands versus the decay of dark energy into matter.
The solutions are characterized by the condition 
$w_\Lambda^{\rm eff}<w_m^{\rm eff}$ for which the late time (fixed point)
solution is
$w_\Lambda^{\rm eff}=w_m^{\rm eff}=-1/3+2q/3$. 

The resulting equilibrium between dark energy and 
matter offers a possible solution to the cosmic coincidence problem. 
For example, the viability of holographic 
models using the future horizon rather than the particle horizon is clearly 
a result of the effective equation of state decreasing with $\Omega _\Lambda$.
However it should be clear that the 
existence of an equilibrium solution is more 
general than the holographic principle in terms of the future horizon.
Detailed quantitative comparison of interacting models with 
present and future observational data 
should be able to further delineate between them.

The interacting models have been generalized to the case 
involving more than two interacting 
components. The resulting evolution equation were shown to 
have a universal form in terms 
of the effective equations of state.

\section*{Acknowledgments}
This work was supported in part by the U.S.
Department of Energy under Grant No.~DE-FG02-91ER40661.

\end{document}